\documentstyle[12pt]{article}
\begin{document}
\newcommand{\be}{\begin{equation}}
\newcommand{\ee}{\end{equation}}

\date{
\setlength{\unitlength}{\baselineskip}
\begin{picture}(0,0)(0,0)
\put(9,19){\makebox(0,0){UUITP-2/93}}
\put(9,18){\makebox(0,0){ITEP-xx/93}}
\put(8,17){\makebox(0,0){hepth/yymmddd}}
\end{picture}}
\title{On Connection between Topological Landau-Ginzburg Gravity
and Integrable Systems.}

\author{A.\ Losev \\{\em Institute of Theoretical and Experimental
 Physics}\\ {\em B.Cheremushkinskaya 25} \\ {\em 117259  Moscow Russia}
\thanks{E-mail address: lossev@nbivax.nbi.dk \&
                        lossev@vxitep.itep.msk.su} \\
I.\ Polyubin \\
{\em Institute of Theoretical Physics} \\
              {\em Thunbergsv\"agen 3 Box 803} \\
               {\em S-751 08 Uppsala Sweden}
\thanks{E-mail address polyub@grotte.teorfys.uu.se \&
          polyub@cpd.landau.free.msk.su}
\thanks{permanent address: L.D.Landau Institute
                     for Theoretical Physics, 117334 Ul. Kosygina 2
                     Moscow Russia.}
 }


\maketitle

\begin{abstract}

  We study flows on the space of topological Landau-Ginzburg theories
coupled to topological gravity. We argue that flows corresponding to
gravitational descendants change the target space from a complex plane to
a punctured complex plane and lead to the motion of punctures.It is shown
that the evolution of the topological theory due to these flows is given by
dispersionless limit of KP hierarchy. We argue that the generating function
of correlators in such theories are equal to the logarithm of the
tau-function of Generalized Kontsevich Model.
\end{abstract}

\newpage
\section{Introduction}
One of the most interesting features of topological matter coupled
to topological gravity\cite{OEW,DW,OV,DVV} is its connection to integrable
systems. Namely, the exponent of the generating function for
topological correlators turns out to be tau-function of the
integrable system. Now we know several possible
ways to explain it:\\
1. Topological theory is in some sense equivalent to the
"physical" theory(conformal matter coupled to 2d gravity)
and the latter can be discretized with the help of matrix
models which expose integrable structures before taking the
 continuum limit \cite{GMMMO} as well as in the double scaling limit
  \cite{Douglas,FKN,DVV,GKM}\\
2. Pure gravity can be described in terms of the
combinatorial model of the moduli space, and this description
naturally arises from the perturbative expansion in the Kontsevich matrix
model ; the matrix integral in this model
turns out to be tau-function of KP hierarchy
\cite{Kon,GKM}\\
3. The so called $\tilde{W}$-constraints can be considered as
symmetries of topological theories;
$\tilde{W}$-algebra naturally acts on tau-function \cite{OG},
 thus the solution to these
constraints is some special tau-function \cite{FKN2,GKM,ASS}\\
4. String field theory for "topological strings" seems to be
a 2d integrable theory \\

We think, that, at present, all these explanations do not give
absolutely satisfactory picture of what happens. That is why
in this paper we
will try to push forward the fifth explanation:\\
5. Observables can be identified with the tangent vectors to the
space of topological theories, thus, each observable leads to
the flow on the space of topological theories. The very existence of the
generating function for the correlators means that these flows commute.

Namely, we consider what goes on in Landau-Ginzburg theories coupled to
topological gravity for the genus zero world-sheet .

 In the Section 2 we summarize results on  flows
corresponding to the primitive observables (lowest times in terms of
KP hierarchy reductions); these flows change superpotential and
observables(due to contact terms). We argue that these contact terms
can be obtained via Verlinde-Verlinde
(VV) mechanism\cite{OV}, i.e. have purely gravitational
interpretation. Different properties of correlators are discussed,
and among them:
 evolution system  and interpretation of the reduced  string
  equation(small phase space at genus zero,
 in the form of Krichever\cite{Kr}) as a dilaton equation.
 The purpose of this section is to fix notations, the only new result is
 the new interpretation of the reduced string equation.

In the Section 3 we argue that flows corresponding to
descendants change the target space of the theory: from a complex plane to
a punctured complex plane. After that change flows correspond to motion of
 punctures. Contact terms
are obtained according the same "rules of the game" as in Section 2;
evolution system in this case is described, properties of correlators
are studied.
We show how dispersionless limit of KP hierarchy appears from the
evolution system.
 Now the full string equation (in the form of Krichever\cite{Kr})
also follows from the dilaton equation.

In the last Section we compare generating functions for correlators
and tau-function of the Generalized Kontsevich Model\cite{GKM}.

In this paper we will be extremely brief in "physical" arguments
(mostly they are given in comments and footnotes); in full detail we will
present them in \cite{lp}.
\section{Simplest case: target space - complex plane}
 \subsection{Primary fields,descendants and 3-point correlator }
 Here we will consider the simplest target space:
complex plane.
Thus, Landau-Ginzburg topological matter (LGM )\cite{Li,Eg,Va} with a complex
plane as a target space
is described by its superpotential $W(X)$
of degree $p$ and a  metric, that is constant in a holomorphic
coordinate $X$ on this plane.\footnote{Due to anomaly in general coordinate
 transformations
on the target space, see\cite{W}, theory can be defined only for metric
whose determinant is a square of the modulus of the top holomorphic form}
The top holomorphic form corresponding to
this metric is equal to $dX$.

The  space of LGM
topological observables form so called chiral ring $C[X]/[dW/dX]$
that is a quotient ring of polynomials $P(X)$
 over the ideal generated by the
derivative of superpotential. Correlators  in LGM in genus zero
are given by:
\be
<P_1 ,\ldots ,P_n>_{W}^{M}=\int_{\Gamma} \frac{P_1(X) \ldots P_n(X)
(dX)^2}{dW}
\ee
Here contour $\Gamma$ goes around infinity in the complex plane.

Superscript $M$ here is very important - it distinguishes correlators in LGM
 from correlators in
  LGM coupled to topological gravity, i.e. Landau-Ginzburg
  gravity (LGG). Later will have no subscript at all.

Appearance of non-covariant term $(dX)^2$ in the correlator on a sphere
illustrates the above-mentioned anomaly in coordinate transformations on
the target space - holomorphic top form appears in power 2 that is the
Euler number of a sphere.

In \cite{los} (see also \cite{Eb}) it was shown \footnote{
The origin of this effect is normal ordering of the operator $X\overline{X}$
- it is similar to appearance of anomalous dimension in conformal theory}
 that the space of local
topological observables in LGG can be embedded in $C[X]$, and
$\sigma_k(\Phi)$ - the
i-th gravitational descendant of the LGM primary field $\Phi$
(by the primary field we call polynomial $\Phi(X)$ of degree less than
 $p-1$  which is the the degree of $W'$)
\footnote{ Primary fields are distinguished
among all polynomials that belong to the same class in the
factor-ring
$C[X]/[dW(X)]$ by the following requirement: acting on a vacuum they
create states with zero energy
\cite{DVV}. If $W$ is a monomial , there is a $U(1)$
 group
that acts on the space of all polynomials, and it is evident that primary
field should have definite charge with respect to this group,so
there is no other choice for primary fields but monomials of order
less than the order of $W$. For an arbitrary $W$ the problem of finding
primaries deserves special investigation } -
can be expressed as follows(see also):
\be
\sigma_i(\Phi)=W'(X)\int^{X} dW(X_i) \ldots
\int^{X_3} dW(X_2) \int^{X_2} \Phi(X_1) dX_1
\ee

In topological gravity correlators are integrals over moduli space of
Riemann surfaces with punctures. In genus zero such moduli space
does not exist for zero, one and two punctures, so we take these
n-point correlators in LGG to be zero for $n<3$. For $n=3$
moduli space is a point and correlator in LGG equals to 3-point
correlator in LGM.
\subsection{4-point  correlators and contact terms}
      It was argued in \cite{los}
 that the four-point correlators with the fourth field being
primary one $<P_1 , P_2 , P_3 , \Phi>_{W}$
describes the derivative of superpotential $W$ plus contact terms:
\begin{eqnarray}
&&<P_1 , P_2 , P_3 , \Phi>_W= \frac{d}{dt}<P_1 , P_2 , P_3>_{W+t\Phi} |_{t=0}
\nonumber \\
&&+<C_W(\Phi , P_1), P_2 , P_3>_W+<C_W(\Phi,P_2), P_1 , P_3>_W
\nonumber \\
&&+<C_W(\Phi,P_3) , P_1 , P_2>_W
\end{eqnarray}
where the contact term of the fields $P$ and $\Phi$ is obtained as
follows. The product of fields (this product naturally appears when these
two fields are placed on the decoupled sphere in Deligne-Mamford
 compactification) is decomposed into a sum of descendant
and a primary fields; primary field part gives no contribution to the
contact term, while descendant contributes due to VV mechanism
\cite{OV}, namely:
\be
P(X) \Phi (X) = W'(X)\int^{X} C_W(P,\Phi)(X_1)dX_1 + \tilde{\Phi}(X)
\; \;  ; deg \tilde{\Phi}<p-1
\ee
 From (4) we see that contact term can be also represented as:
\be
C_W(P,\Phi)=\frac{d}{dX} (\frac{P\Phi}{W'})_{+}
\ee
Here subscript "+" stands for non-negative powers of $X$
 in expansion at infinity.

One can check that 4-point  correlator (4) is symmetric under permutations
of fields (as it should be).

\subsection{Formalized definition of $n$-point  correlator}
By induction one can define the $n$-point  correlators according to the
 following four rules:

{\bf 1.} The $n$-point  correlator is a symmetric linear functional
 on $C[X]^{\otimes n}$,
denoted as $<P_1 ,\ldots , P_n>_W$

{\bf 2.}This functional is equal to zero for $n < 3$ and is given by the
Grothendieck residue (1) for $n=3$.

{\bf 3.}If all fields are descendants \footnote
{if all fields are descendants then the
complex dimension of form to be integrated
 over the moduli space is at least $n$
 while the dimension of moduli space is
 $n-3$ }, this functional equals to zero
\be
<\sigma_{I_1}(\Phi_1), \ldots , \sigma_{I_n}(\Phi_n)>_W=0,\; \; if \; \;
I_i  \neq 0\, \, for\, \, all\, \, i
\ee

{\bf 4.} If one of the fields is primary and $n>3$ then multi-point
 correlators
 obey the following
recursion relations\footnote{these relations correspond to integration
over moduli corresponding to the position of the insertion of the field
$\Phi$, i.e. to the "motion" of the point
where $\Phi$ is inserted while all other points are keeping fixed}
(these recursion relations should not be mixed with recursion
relations, connecting correlator of $\sigma_i$ and $\sigma_{i-1}$):
\be
<P_1 ,\ldots , P_n ,\Phi>_W=
\frac{d}{dt}<P_1 ,\ldots , P_n >_{W+t\Phi}|_{t=0} + \sum_{i=1}^{n}
<P_1 , \ldots , C_W(P_i,\Phi) ,\ldots , P_n >_W
\ee
and bilinear multiplication(contact term)
 $C_W: C[X]^{\otimes 2} \rightarrow C[X]$
is given in (5).
We will call the field $\Phi$ that enters in recursion relation (7)
a "moving" field.

One can check that this definition of multi-point  correlator is
self-consistent.

\subsection{Properties of multi-point  correlators}
 From the definition one can check the following properties of
multi-point correlators.

{\bf 1. Puncture equation.}

 Let us denote as $1_P$ polynomial that equals
to $1$ to distinguish it from all other "one's" that can appear in
the algebra; then:
\be
<\sigma_{I_1} (\Phi_1) , \ldots , \sigma_{I_n}(\Phi_n), 1_P>_W=
\sum_{i=1}^{n}
<\sigma_{I_1}(\Phi_1) , \ldots ,
\sigma_{I_i-1}(\Phi_i) , \ldots  ,  \sigma_{I_n}(\Phi_n) >_W
\ee
here we accept $\sigma_{-1}(\Phi)=0$

{\bf 2.Dilaton equation.}

Using standard terminology we will call the field
\be
\sigma_{1}(1_P)=XW'(X)
\ee
the dilaton.
Then recurrently one can show that
\be
<\sigma_{1}(1_P), P_1 ,\ldots , P_n >_W=
(n-2)<P_1 ,\ldots ,P_n>
\ee
This equation(called the dilaton equation)
 is expected on general grounds from topological gravity ($2-n$ is an
 Euler characteristic of sphere with $n$ deleted points)

 {\bf 3.Factorization equation.}

 If $\sum_{i=1}^{n}I_i=n-3$ then
 \be
 <\sigma_{I_1}(\Phi_1), \ldots ,\sigma_{I_n}(\Phi_n)>_W=
 \frac{(n-3)!}{\prod_{i=1}^{n} I_i !}
 \int \frac{\Phi_1\cdot \ldots \cdot \Phi_n (dX)^2}{dW}
 \ee
 Note that the second factor in (11) is nothing but the $n$-point correlator
 in topological matter, and equation (11) is also expected on general grounds
  \cite{DVV,DW,OEW} from topological gravity coupled to topological matter.

 \subsection{Introduction of times on small phase space}
 Let us consider the generating function for all multi-point correlators.
 We will define a correlator of fields $P_1, \ldots P_n$ in
 the presence of a formal exponent of a field  $P$
 as a generating function for multipoint correlators :
 \begin{eqnarray}
 <P_1 ,\ldots , P_n ; \exp (P)>_W &=&
<P_1 ,\ldots , P_n >_W + <P_1 ,\ldots , P_n , P>_W+ \nonumber \\
 \frac{1}{2}<P_1 ,\ldots, P_n ,P, P>_W&+&\ldots+
 \frac{1}{k!}<P_1 ,\ldots , P_n , P , P , \ldots , P>_W+\ldots
 \end{eqnarray}
 Then, given a basis of primary fields $\Phi_a$ and a set of parameters
  $t_a$ (these parameters are called coordinates on the small phase space)
 we  define a $t$-dependent multi-point correlators:
  \be
<P_1 , \ldots , P_n >_W(t)= <P_1 , \ldots , P_n  ;\exp(\sum_{a=1}^{p-1} t_a
\Phi_a)>_W
  \ee
  From the definition of multi-point correlators it follows that for
  $n>2 $ one can represent the t-dependent multipoint correlators in
  terms of ordinary multipoint correlators as:
  \be
  <P_1 , \ldots , P_n >_W (t)=<P_1(t) , \ldots , P_n(t) >_{W(t)}
  \ee
  Where $W(t)$ and $P(t)$ are solutions to the following system of nonlinear
  differential equations:
  \be
  \frac{\partial}{\partial t_b}\Phi_a(t)=C_{W(t)}(\Phi_a(t),\Phi_b(t))
  \ee
\be
\frac{\partial}{\partial t_b}W(t)=\Phi_b(t)
  \ee
\be
\frac{\partial}{\partial t_b}P_i(t)=C_{W(t)}(\Phi_b(t),P_i(t))
  \ee

  In this derivation it is important that contact term between primary
  fields gives a primary field that can also be used in recursion
  relations (7).

  From the physical setting we expect that system is integrable, one can
  explicitly check that it is so.

  \subsection{Flat times}
  Here we will explain how flat times on the space of potentials
  appear in the evolution system (15-17) \footnote{Here we closely follow
   derivation presented in \cite{DVV}}. Using recursion relations(7)
    with $1_P$ as a moving
  field one can see that for all positive $n$:
  \be
  < \Phi_a , \Phi_b , \Phi_1 , \ldots , \Phi_n , 1_P>_W=0
  \ee
  This follows from two facts:\\
  1. the contact term of $1_P$ with primary field is zero\\
  2. adding constant to $W$  does not change $dW$ and it is $dW$
  that appears in  all other recursion relations.

  Using the symmetry of the multipoint correlators we see that (14) means
  \be
  \eta_{ab}(t)=<\Phi_a(t), \Phi_b(t) , 1_P>_{W(t)}
   =<\Phi_a , \Phi_b , 1_P>_{W}
  \ee
  here we used that $1_P(t)=1_P$

   Primitive fields naturally form a tangent space to the space
   $A_p$ of all
   superpotentials of degree $p$ moduli
   constant shifts and dilatations . Together with (1) it means that $\eta$
   is a metric on this space in coordinates $t$ , and (19) means that it is
   flat, that is the set of times $\{ t_a \}$ forms flat coordinates on $A_p$.

  \subsection{Relation between $P(t)$ and $W(t)$}
  Evolution system can be partly integrated. Namely, one can show that
  \be
  \partial_X ((W(X,0)^{j/p})_{+} (t)=  \partial_X ((W(X,t))^{j/p})_{+}
  \ee
  Here the left hand side is understood as a polynomial P(t) that
  evolves due to system (15-17) and that equals to
  $  \partial_X ((W(X,0))^{j/p})_{+} $ when all times are set to zero;
  $W(X,t)$  is a superpotential that equals to $W(X)$ when all times are zero.

  Since evolution of primitive polynomials $\Phi$
   is a particular case of evolution
  of general polynomials $P(t)$ (it corresponds to $j<p $ in (20)), one
  can substitute $\Phi_a(t)$ from (20) into flat times equation (19)
  and see that this equation is fulfilled.

\subsection{Solution to the evolution system; its
interpretation
  as a consequence of the dilaton equation}
  Let as take for monomial superpotential $W(X,0)=X^{p}$
a basis in the space of polynomials:
  \be
  P_j(X,0)= j X^{j-1} , \; \; thus,
    \; P_j(X,t)=\partial_{X}(W(X,t)^{j/p})_+
  \ee

  Then one can see that  for $t=0$ $\frac{p}{p+1} P_{p+1}(X,0)$ is a dilaton.
  Using dilaton equation one can easily check that for $k>2$
  \begin{eqnarray}
&&  <\frac{p}{p+1}P_{p+1}(0), P_{a_1}(X,0),\ldots , P_{a_k}(X,0) ; \exp (
\sum_{i=1}^{p-1}t_i P_i(X,0) )>_{W(X,0)}=
\nonumber \\
&&  (k-2) <P_{a_1}(X,t) , \ldots , P_{a_k}(X,t) ;
 \exp (\sum_{i=1}^{p-1}t_i P_i(X,0) )>_{W(X,0)}+
\nonumber \\
&&  <(\sum_{i=1}^{p-1} t_i P_i(X,0)), P_{a_1}(X,0),\ldots , P_{a_k}(X,0);
   \exp (\sum_{i=1}^{p-1}t_i P_i(X,0) )>_{W(X,0)}
  \end{eqnarray}
  Putting the second term to the left-hand side and using evolution
system we get:
  \begin{eqnarray}
   &&<[\frac{p}{p+1}P_{p+1}(X,t)-(\sum_{i=1}^{p-1} t_i P_i(X,t))] ,
    P_{a_1}(X,t) , \ldots , P_{a_k}(X,t)>_{W(X,t)}= \nonumber \\
  &&(k-2) <P_{a_1}(X,t) ,\ldots , P_{a_k}(X,t)>_{W(X,t)}
  \end{eqnarray}
  Thus the term in the square brackets behaves as a dilaton for
  superpotential $W(t)$, but we already know (9) how it looks like,
  so we obtain;
  \be
  \frac{p}{p+1}P_{p+1}(t)-(\sum_{i} t_i P_i(X,t))=X\partial_X W(X,t)
  \ee
  This equation\footnote{one can call it reduced
  string equation,in this form it appeared in \cite{Kr}}
   (together with (21)) solves the evolution system.

\section{Higher times and punctured complex plane }
This section is a result of trying to understand results
of \cite{Kr} (see also \cite{du}) in physical terms.
\subsection{Motivation}
One may ask what happens if we consider generating function not only
for primaries but also for descendants. The answer is that correlators in
the presence of such generating
function can be interpreted as
correlators in LGG theory on a punctured complex plane.
Namely, descendants are of the form $W'R(X)$ and naively could be
considered as terms that come from the Ward identities
connected to reparametrization of the
complex plane, i.e. naively one would expect something like:
\be
<P_1 , \ldots , P_n , W'R >_W + <P_1'R(X) ,\ldots , P_n>_W + \ldots +
  <P_1 ,\ldots , P_n'R>_W=0
  \ee
Putting all the terms except first to the right-hand side one could try to
  consider $W'R$ as a generator of reparametrization of  the
target space :  naively such a reparametrization results into "classical
contact terms"(we call them "classical" because they origin
 from the very classical
consideration of changing variables in the functional
integral)\\
  But it is not the full story,because:\\
  1.there should be nontrivial "gravitational" contact terms\\
  2.LG theory (as a
  sigma-model  of type B!) is described not only by
  superpotential but also by a holomorphic top form (that we previously
  considered to be equal to $dX$). After reparametrization with
  parameter $t$
  this top form (that we will denote as $dQ$)
  becomes
\be
 dQ= dX(1+tR'(X)),
\ee
so $dQ$  degenerates  at zeros of $1+tR'(X)$.
  So we inevitably have to consider LG theories with such degenerated
  form.
  We will discuss this theory in the same manner as the
  ordinary LG theory.
  \subsection{Observables,descendants,primitive fields and LGM correlator
  in a theory on a punctured complex plane}
  Theory on a punctured complex plane is defined by the superpotential
  $W$ (that is a holomorphic function with pole only at
   infinity) and the holomorphic top form $dQ$ that has zeros
   only at the punctures and a pole at infinity. Moreover, we
   demand that at puncture(at zero of $dQ$)  the $dW$ should be
   nondegenerate (nonzero). \footnote{These conditions
correspond to LG theory with the F-term defined by a
holomorphic function $W$ and D-term defined by a Kahler
prepotential $Q\overline{Q}$; $Q$ is an integral of the form
$dQ$}

  {\bf Observables}

The space of observables is given by functions
  of the form $P(X)/Q'(X)$. One can show that $W'/Q'$ is on shell
  trivial obsevable that is why observables in LGM are $P/Q'$ where
 $P$ belongs to
  the same factor-ring $C[X]/W'$.

  {\bf Descendants}

 Descendants have the form
  \be
\sigma_{1}^{W,Q}(P/Q')=dW/dQ\int P/Q'dQ= \sigma_{1}^{W,X}(P)/Q'
\ee
{\bf Primitive elements}

   Primitive
  elements are $\Phi/Q'$, where degree of $\Phi$ is
   less than degree of superpotential. Note that
   the simple puncture  $1_P= Q'/Q'$
  can be  not a primitive element, and in all cases is not an
  element of lowest degree any more! But  the dilaton is
   still $\sigma(1_P)$!

{\bf Correlators in LGM}

  Correlators in the LG topological matter on the punctured
plane look very similar to those for a smooth plane:
the difference is in using $dQ^2$ instead of $dX^2$:
\be
<P_1/Q', \ldots, P_n/Q'>_{W,Q}^{M}
=\int_{\Gamma} \frac{P_1(X)/Q' \ldots P_n(X)/Q' (dQ)^2}{dW}
\ee
Here the contour of integration $\Gamma$ separates zeros of
$dW$ from infinity and punctures (equally one can say that
this contour goes around the punctures).
For example, the three point correlator is given by
(this formula already appeared in \cite{Kr}
\be
<P_1/Q', P_2/Q', P_3/Q'>_{W,Q}^{M}=\int_{\Gamma} \frac{P_1(X) P_2(X) P_3(X)
 (dX)^3}{dW dQ}
\ee
and the metric $\eta$ - two-point correlator:
\be
<P_1/Q', P_2/Q'>_{W,Q}^{M}=\int_{\Gamma} \frac{P_1(X) P_2(X) (dX)^2}{dW}
\ee
\subsection{Recursion relations}
To treat the field $P/Q'$ as a "moving" field we have to
decompose it into a sum of two fields: the field
$S$ that is a
polynomial of degree less than the degree of superpotential
and the field $RW'/Q'$:
\be
P=Q'S + W'R
\ee
The decomposition (31) exists and is unique if and only
if polynomials $Q'$ and $W'$ are mutually simple, i.e. do not
have common zeros.

{\bf S-recursion relations}

The S-recursion relations correspond to the "motion " of field
$S$ and are not very different from recursion relations for
unpunctured case.
The motion of the fields  $S$ corresponds to the infinitesimal
(perturbative) shift in superpotential $W \rightarrow
W+tS $, while contact term
$C_{W,Q}(P_1/Q',S) $ of $S$ with some other field $P_1/Q'$ is
given by  VV rule and can be found from:

\be
SP_1/Q'=\Phi/Q' + W'/Q' \int C_{W,Q}(P_1/Q',S) dQ
\ee
This contact term can be easily calculated,see (5):
\be
 C_{W,Q}(P_1/Q',S)=1/Q'  C_{W,X}(P_1,S)=
1/Q' (P_1S/W')_{+}'
\ee
Thus, we get the S-recursion relations in the LGG:for $n>2$
\begin{eqnarray}
&&<P_1/Q', \ldots P_n/Q', S>_{W,Q}=
\frac{d}{dt}<P_1/Q', \ldots , P_n/Q'  >_{W+tS,Q}
\nonumber \\
&& + \sum_{i=1}^{n}
<P_1/Q',  \ldots , C_{W,Q}(P_i/Q',S ), \ldots , P_n/Q'  >_{W,Q}
\end{eqnarray}

{\bf R-recursion relations}

R-recursion relations, that correspond to the "motion" of the
field $RW'/Q'$ arise from the Ward identities connected with
the reparametrization of correlators:
\be
X \rightarrow X + tR(X)/Q'(X)
\ee
Due to this reparametrization in $<P_1, \ldots , P_n>_{W,Q}$
\begin{eqnarray}
W & \rightarrow & W + tW'R/Q'\nonumber \\
Q & \rightarrow & Q + tR  \nonumber \\
P_i/Q' & \rightarrow & P_i/Q' + t(P_i/Q')'R/Q'= P_i/Q' -
t C_{W,Q}^{cl} (P_i/Q',RW'/Q')
\end{eqnarray}
The change of $P_i$ due to reparametrization will be
interpreted below as minus the "classical" contact term
$C^{cl}$ . But we know that
there is also a "topological gravity" contact term $C^{tg}$
(that we once again obtain due to VV rule):
\be
P_i R W'/(Q')^2= W'/Q' \int C_{W,Q}^{tg}(P_i /Q',RW'/Q') dQ
\ee
thus
\be
 C_{W,Q}^{tg}(P_i /Q',RW'/Q')  = (P_i R/Q')'/Q'
\ee
Now we are ready to use Ward identities; they state (for $n>2$)
\begin{eqnarray}
&&<P_1/Q', \ldots , P_n/Q', RW'/Q'>_{W,Q}=\frac{d}{dt}
(<P_1/Q', \ldots, P_n/Q'>_{W,Q-tR})|_{t=0}+\nonumber \\
&&\sum_{i=1}^{n}<P_1/Q', \ldots ,
C_{W,Q}^{tot}(P_i /Q',RW'/Q')  ,  \ldots , P_n/Q'>_{W,Q}
\end{eqnarray}
where the total contact term is given by:
\begin{eqnarray}
&&C_{W,Q}^{tot}(P_i /Q',RW'/Q') =
\nonumber \\
&& C_{W,Q}^{tg}(P_i /Q',RW'/Q') +
C_{W,Q}^{cl}(P_i /Q',RW'/Q') = P_i R'/(Q')^2
\end{eqnarray}
Thus, the R-recursion relation takes the following simple
form:
\be
<P_1/Q', \ldots , P_n/Q', RW'/Q'>_{W,Q}=\frac{d}{dt}
<P_1/(Q-tR)', \ldots , P_n/(Q-tR)'>_{W,Q-tR}|_{t=0}
\ee
With the help of S and R recursion relations (together with
the three point correlator), all correlators are defined.
Moreover, one can recursively prove
(expected) generalized dilaton and
factorization formulas.They look as follows.
\subsection{Dilaton equation and factorization formula on a
punctured plane}
\be
<W'Q/Q' , P_1/Q',  \ldots , P_n/Q'  >_{W,Q}=
(n-2)<P_1/Q' , \ldots , P_n/Q' >_{W,Q}
\ee
Note, that on a punctured plane dilaton is
the descendant of the puncture $1_P$, i.e. $W'Q/Q' $, see (28).

 If $\sum_{i=1}^{n}I_i=n-3$ then
 \be
 <\sigma_{I_1}(\Phi_1/Q'), \ldots ,
\sigma_{I_n}(\Phi_n/Q')>_{W,Q}=
 \frac{(n-3)!}{\prod_{i=1}^{n} I_i !}
 \int \frac{\Phi_1\cdot \ldots \cdot \Phi_n (dX)^2}{dW
(Q')^{n-2}}
 \ee
 \subsection{Introduction of all times and evolution system}
 From S and R recursive equations we see that  times,
corresponding to descendants of primary
 observables in generating exponent  lead to evolution of all
ingredients of the theory: insertions, superpotential and the
top form:
  \begin{eqnarray}
&& <P_1/Q',  \ldots , P_n/Q' ; \exp(\sum_{k=1}^{\infty}t_k
P_k/Q' )>_{W,Q} = \nonumber\\
&& <P_1(t)/Q'(t), \ldots , P_n(t)/Q'(t)  >_{W(t),Q(t)}
  \end{eqnarray}
 If we define polynomials $R_j(t)$ and $S_j(t)$ as a R and S parts
of the polynomial $P_j(t)$:
\be
P_j(X,t)=W'(X,t)R_j(t)+Q'(X,t)S_j(t)
\ee
then
\begin{eqnarray}
\frac{\partial}{\partial t_j} P_i(t)& =&
C_{W,X}(P_i(t),S_j(t))= (P_i(t) S_j(t)/W'(t))'_{+}  \\
\frac{\partial}{\partial t_j} W(t)&=& S_j(t)  \\
\frac{\partial}{\partial t_j} Q(t)&=& -R_j(t)
\end{eqnarray}
Below to save space we will write $\partial_i$ instead of
$\frac{\partial}{\partial t_i}$.

This system can be easily partly integrated, namely,
\be
P_j(t)= (W(t)^{j/p})'_{+}
\ee
 satisfy this system.

In other terms relation (45) can be rewritten as
\be
P_j(X,t)=\partial_j W(X,t) \partial_X Q(X,t)-\partial_j Q(X,t)
\partial_X W(X,t)
\ee
This relation can be reinterpreted if we
 introduce the Poisson bracket, connected with the i-th time as
an antisymmetric bilinear function on any two functions of
$X$ and $t_i$ (let us call them $T_1 $
and $T_2$ ). This i-th bracket is given by the formula
\be
\{ T_1(X,t_i) , T_2(X,t_i) \}_i=\partial_i T_1(X,t_i) \partial_X T_2(X,t_i)
 - \partial_i T_2(X,t_i) \partial_X T_1(X,t_i)
\ee
Namely, the decomposition of the observable on shifts in superpotential
and reparametrization of the target space (45) is equivalent to:
\be
\{ W, Q \}_i = P_i
\ee
If we take as $P_i$ the identity , then due to evolution it remains
identity, so we have the specialization of (52) in the form:
\be
\{ W, Q \}_1 = 1
\ee
\subsection{Evolution system (46-48) as a dispersionless limit of
the reduced KP}
In what follows we will use the following simple:\\
{\bf Lemma}\\
Let us for a polynomial $P$ denote by $[P]$ its part that has
power less then the power of $W'$, then:
\be
[ \partial_i W P_j]=[\partial_j W P_i]
\ee

This lemma can be proven by multiplication of equation (50) on
$\partial_i W$
and observation that up to terms proportional to $W'$ it is symmetrical
in $i$ and $j$.

{\bf Statement}\\

If $W(X,t)$ and $P_m(X,t)$ are solutions of the evolution system (46-48),
 then
\be
\{ W(X,t), \int^X P_j(X_1,t)dX_1 \}_i = [P_i \partial_j W ]
\ee
Proof.
\be
\partial_i W P_j= [\partial_i W P_j ] +
\partial_X W \int^X C_{W,X_1}(\partial_i W,
P_j)dX_1=[\partial_i W P_j] + \partial_X W \partial_i \int^X P_j dX_1
\ee
Application of Lemma (54) ends the proof.

Specializing statement to the case $P_i(X,t)=1$ we get
\be
\{ W(X,t), \int^X P_j(X_1,t)dX_1 \}_1 =  \partial_j W(X,t)
\ee
Further specializing to the case when at $t=0$ superpotential
and all observables are monomials, we obtain
\be
\{ W(X,t),  (W(X,t)^{j/p})_{+} \}_1 =  \partial_j W(X,t)
\ee
Equation that we get means that evolution (46-48) of $W$ coincides
with the evolution of $W$ due to dispersionless reduction of KP!

Next we can interpret the evolution of $Q$. Already from (53) we
observe that evolution preserves Poisson bracket between $W$ and $Q$,
so the evolution is expected to be a simplectomorphism. Now we will
show that it really happens. Really, taking $\partial_i Q P_j$ and
making with it manipulations like in proof of the lemma and in
proof of the previous statement one can easily show the following

{\bf Statement}
\be
 \{ Q(X,t), \int^X P_j(X_1,t)dX_1 \}_i =
  P_i \partial_j Q - \int^X C_{W,X_1}(\partial_j W P_i)dX_1
\ee
Specially, for $P_i=1$ and in the homogeneous case
 we get the desired relation
\be
\{ Q(X,t), (W(X,t)^{j/p})_{+}  \}_1 =  \partial_j Q(X,t)
\ee
that means that evolution system is equivalent to symplectomorphisms
(with respect to Poisson bracket $\{,\}_1 $ ) generated by
$(W(X,t)^{j/p})_{+}$.

\subsection{Solution of the full evolution system from a dilaton equation}
Reasoning exactly like in (24) one can show that if for zero
times $Q=X$ and $W=X^p$, and as a basis in the space of
observables we take monomials, then for arbitrary times
the following observable
\be
 [\frac{p}{p+1}(W(t)^{p+1/p})'_{+}-\sum_{j=1}^{\infty} t_j
(W(t)^{j/p})'_{+}]/Q'
\ee
 satisfies dilaton equation.
Comparing this object with the dilaton that is a descendant of
a puncture(42) , we get the
Krichever form of the string equation in genus zero \cite{Kr}:
\be
\frac{p}{p+1} (W(t)^{p+1/p})'_{+}-\sum_{j=1}^{\infty} t_j
(W(t)^{j/p})'_{+})=QW'
\ee
that determines $W$ and $Q$ in terms of times $t$.
\section{Generating function for correlators and $\tau$ function of the
Generalized Kontsevich Model}
In this section we will argue that
\be
\log Z_{LGG}(t,W)=
\sum_{q=0}^{\infty}<\exp(\sum_{k=1}^{\infty}t_k k X^{k-1})>_{W(X)}^{(q)}
= \log Z_{GKM}(t,V)
\ee
where \\
$<>_{W}^{(q)}$ stands for the
correlator in genus $q$ of the worldsheet in topological LG gravity with
superpotential $W$ \\
$Z_{GKM}(t,V)$ is the partition function of the Generalized Kontsevich
Model (GKM) \cite{GKM}, and $V'=W$. This GKM partition function
$Z_{GKM}$ is a $\tau$-function
of the KP hierarchy.
This $\tau$ function corresponds to $W$-reduction, i.e.
for a $L$ operator of the hierarchy ,$W(L)$ is a differential operator.
Moreover, this $\tau$-function satisfies so called $L_{-1}$
constraint.

Our argument would be two-fold.
We begin by showing that the genus zero contribution to $Z_{LGG}$
satisfies the same equation (71) as $\tau$-function of dispersionless KP.

Our second argument is connection between $Z_{LGG}$
for different superpotentials $W$ of the same degree. We will see that
if $t$ are some times on the small phase space, and $T$ are times on the
full phase space,then for some matrix $c(t)$
$$\log Z_{LGG}(t+T,W)= \log Z_{LGG}(c(t)T,W(t)) +A(t,T), $$
 additional term $A$ considered as a function of $T$
 is a polynomial of second degree. This term is coming only from genus zero
  of the worldsheet
and means simply that there is no moduli space for the sphere with
zero, one or two punctures. The same relation between partition
functions in GKM was proven in \cite{Kha}.
\subsection{Generating function for genus
 zero correlators as a $W$-reduced dispersionless $\tau$ - function }
To have KP hierarhy we need an $L$ operator, in dispersionless KP this
operator becomes a function. So
we define $L(X,t)$ as such a formal seria in $X^{-1}$, that behaves as
$X$ at infinity in the $X$-complex plane, and that satisfies:
\be
W(L(X,t),0)=W(X,t)
\ee
{}From equation (57) it is obvious that evolution of $L$
is given by the Poisson bracket with Hamiltonians:
\be
\{ L(X,t), \int^X P_j(X_1,t)dX_1 \}_1 =  \partial_j L(X,t)
\ee
where (due to definition of times in (63))
\be
P_j(X,t)=\sum_{k=1}^{\infty} \tilde{c}_{jk}\partial_X(W(X,t)^{k/p})_{+} ,
\ee
coefficients $\tilde{c}_{jk}$ are $t$ and $X$ independent, they are defined
by $W(X,0)$ as follows:
\be
j X^{j-1}=\sum_{k=1}^{\infty} \tilde{c}_{jk}\partial_X(W(X,0)^{k/p})_{+}
\ee

{\bf Lemma}\\
Let $\partial$ denote some partial derivative along times or  $X$ ,then
\be
\partial (L^j )_{+}=\partial \int^{X} P_j(X_1,t)dX_1
\ee

Proof.\\
The right hand side of (68) equals to
\begin{eqnarray}
&&\sum_{k=1}^{\infty} \tilde{c}_{jk} k/p (W(L(X,t),0)^{k/p-1}
\frac{\partial W(L,0)}{\partial L} \partial L(X,t))_{+}=
\nonumber \\
&&\sum_{k=1}^{\infty} \tilde{c}_{jk} k/p ((W(L(X,t),0)^{k/p-1}
\frac{\partial W(L,0)}{\partial L})_{+,L} \partial L(X,t))_{+}=
 \partial (L^j)_{+}
\end{eqnarray}
Here "$+,L$ " stands for the positive powers in $L$ expansion. The first
equality in (69) is valid because  the negative part in $L$
expansion times $\partial L$
gives zero contribution to the positive part in $X$ expansion of the product.

{}From (65) and (68) we see that $L$ satisfies dispersionless analog of KP,
more exactly its $W(0)$ reduction. Namely
\be
\{ L(X,t), (L(x,t)^j)_+  \}_1= \partial_j L(x,t)
\ee

Now we are ready to prove a statement that shows that partition function of
LGG satisfies the same equation as the tau-function, i.e. :

{\bf Statement}
\be
\partial_j \partial_1\log Z_{LGG}^{0}(t,W(X,0))=
\partial_j \partial_1<\exp(\sum_{k=1}^{\infty}t_k k X^{k-1})>_{W(X,0)}=
\int_{\Gamma} L^{j} dX
\ee

Proof.
\be
\partial_i\partial_j\partial_1 \log Z_{LGG}^{0}(t,W(X,0))=
\int_{\Gamma}\frac{P_i(X,t) P_j(X,t)dX}{W' Q'}=
\int_{\Gamma}\frac{\partial_i W(X,t) P_j(X,t)dX}{W'}
\ee
In the second equality  we used decomposition (50)
 and the fact that contour $\Gamma$ goes around
zeros of $W'$. Using
Lemma (68) we can rewrite the last expression as:
\be
\int_{\Gamma}\frac{\partial_L W(L,0)
\partial_i L (j L^{j-1} \partial_X L )_{+}}{\partial_L W(L,0)
\partial_X L} =
\partial_i \int_{\Gamma} L^j
\ee
In the numerator of (73) operation $()_+$
can be omitted because negative part of expansion gives zero
contribution to the integral.

Thus, up to the constant  statement (71) is proven. Since when all times are
equal to zero the left hand side of the statement is zero (there is no moduli
space for a sphere with 2 punctures), and the right hand side is also zero
(when all times are zero , $L=X$), this constant is zero.

\subsection{Relation between $\log Z_{LGG}(T,W)$ for different superpotentials
  $W$}
  Consider $\log Z_{LGG}(t+T,W(X,0))$ for monomial potential $W(X,0)$ and
for some parameters $t_k$, $k<p$.
 (Note, that here we are dealing
  not with the spherical contribution but with the full
expression). Then due to properties of the formal exponent
\begin{eqnarray}
&&\log Z_{LGG}(t+T,W(X,0))=\sum_{q=0}^{\infty}
 <\exp \sum_{k=1}^{\infty}(t_k+T_k)k X^{k-1}>_{W}^{(q)}= \nonumber \\
&&\sum_{q=0}^{\infty} <\exp \sum_{k=1}^{\infty} T_k k X^{k-1};\exp
 \sum_{k=1}^{p-1}
t_k k X^{k-1}>_{W}^{(q)}
\end{eqnarray}
Naively one could think that due to evolution with respect to times $t$
this expression equals to $\log Z_{LGG}(\sum_{j}c_{kj}(t) T_j,W(X,t))$
where matrix $c(t)$ is given by:
\be
 \partial_X (W(X,t)^{k/p})_{+}=\sum_{j} \tilde{c}_{kj}(t) j X^{j-1}
\; , \;  c_{kj}=\tilde{c}_{jk}
\ee
This is not quite true, since one can apply evolution equation
only for these terms in
the last expression in (74) that are at least third degree
  polynomials in $T$ . Terms that are polynomials of lower
   degree give nonzero contribution to
$\log Z(t+T;W(0))$ but give zero contribution to $\log Z(cT,W(t))$
because there is no moduli space on the sphere with less than three punctures
on it. Taking this into account we get the final\\
{\bf Statement}
\begin{eqnarray}
&&\log Z_{LGG}(\sum_{j}c_{kj}(t) T_j,W(X,t))=
\log Z_{LGG}(t+T,W(X,0))- \nonumber \\
&&(1 + \sum_{k=1}^{\infty}T_k \frac{\partial}{\partial \theta_k}
+\frac{1}{2}\sum_{j,k=1}^{\infty}T_j T_k
\frac{\partial}{\partial \theta_j} \frac{\partial}{\partial \theta_k})
\log Z_{LGG}^{0}(t+\theta,W(X,0)) |_{\theta = 0}
\end{eqnarray}
Exactly this relation between $\log Z_{GKM}(T,V)$ for different potentials
  was found in \cite{Kha}.\\

 \vspace{1cm}

{\bf Acknowledgements:}

 We are deeply indebted to
A. Gerasimov, V. Fock, I. Krichever, S. Kharchev, A. Marshakov, A. Mironov,
A. Morozov, N. Nekrasov, A. Rosly for fruitful  discussions.
We are grateful to Institute of Theoretical Physics at Uppsala University
and to prof.
 A.Niemi for providing a stimulating atmosphere and support.
 A.L. is greatful for warm hospitality to the Niels Bohr Institute where
 this paper has been completed.

\end{document}